\documentclass[aps,prd,reprint,a4paper,floatfix,nofootinbib,superscriptaddress]{revtex4-2}

\usepackage[T1]{fontenc}
\usepackage{lmodern}
\usepackage{amsmath,amssymb,amsfonts,bm,mathrsfs}
\usepackage{graphicx}
\usepackage{booktabs}
\usepackage{enumitem}
\usepackage{xcolor}
\usepackage{comment}
\usepackage[colorlinks=true,linkcolor=red,citecolor=blue,urlcolor=blue]{hyperref}

\ifdefined\pdfsuppresswarningpagegroup
\fi
\newcommand{\Pk}{\mathcal{P}_{\zeta}}
\newcommand{\kMpc}{\,\mathrm{Mpc}^{-1}}
\newcommand{\OmGW}{\Omega_{\rm GW}}
\newcommand{\Omchi}{\Omega_{\chi}}
\newcommand{\fzero}{f_0}
\newcommand{\Az}{A_{\zeta}}
\newcommand{\Rthree}{\mathcal{R}_3}
\newcommand{\Mthree}{\mathcal{M}_3}
\newcommand{\Achi}{\mathcal{A}_{\chi}}
\newcommand{\Mpl}{M_{\rm Pl}}
\newcommand{\Neff}{N_{\rm eff}}

\newcommand{\dd}{\mathrm{d}}

\newcommand{\ee}{\mathrm{e}}

\begin{document}

\title{Conformal dark matter and MHz gravitational waves}

\author{Imtiaz Khan}
\email{ikhanphys1993@gmail.com}
\affiliation{Department of Physics, Zhejiang Normal University, Jinhua, Zhejiang 321004, China}
\affiliation{Research Center of Astrophysics and Cosmology, Khazar University, Baku, AZ1096, 41 Mehseti Street, Azerbaijan}

\author{Niamat Ullah}
\email{Niamat.ullah@buitms.edu.pk}
\affiliation{Department of Physics, Balochistan University of Information Technology, Engineering and Management Sciences (BUITEMS), Quetta, Balochistan, Pakistan.}

\author{Salvatore Capozziello}
\email{capozziello@na.infn.it}
\affiliation{Dipartimento di Fisica ``E. Pancini", Universit\`a di Napoli ``Federico II", Complesso Universitario di Monte Sant’ Angelo, Edificio G, Via Cinthia, I-80126, Napoli, Italy,}
\affiliation{Istituto Nazionale di Fisica Nucleare (INFN), sez. di Napoli, Via Cinthia 9, I-80126 Napoli, Italy,}
\affiliation{Scuola Superiore Meridionale, Vi Mezzocannone 4, I-80134, Napoli, Italy.}

\author{G. Mustafa}
\email{gmustafa3828@gmail.com}
\affiliation{Department of Physics, Zhejiang Normal University, Jinhua, Zhejiang 321004, China}

\author{Farruh~Atamurotov}
\email{atamurotov@yahoo.com}
\affiliation{Inha University in Tashkent, Ziyolilar 9, Tashkent 100170, Uzbekistan}

\begin{abstract}
A localized enhancement of the primordial curvature spectrum can leave two distinct relics: gravitationally produced conformal-fermion dark matter and a scalar-induced stochastic gravitational-wave background. We show that the dark-matter abundance fixes the scalar normalization through the cubic moment of the curvature spectrum, while the induced tensor signal probes its quadratic convolution. This closes the usual normalization freedom in scalar-induced gravitational-wave templates and ties the MHz signal directly to the dark-matter mass, peak scale, and spectral width. A concrete single-field realization demonstrates this mechanism: a Mukhanov--Sasaki evolution produces a broad MHz background that sits safely below the Gaussian primordial-black-hole threshold. In this construction, a null high-frequency search becomes a lower bound on the conformal-fermion mass, while a detection has to reproduce the relic abundance, peak frequency, amplitude, and width from one primordial feature. The result is a testable link between small-scale inflationary structure, superheavy dark matter, and laboratory MHz gravitational-wave searches
\end{abstract}

\maketitle

\section{Introduction}

CMB and large-scale-structure data fix the primordial spectrum near \(k\sim0.05\,\kMpc\), including its amplitude, tilt, and running \citep{Aghanim:2018eyx,Ijaz:2026ear}. Many orders of magnitude in comoving wavenumber remain unmeasured. Those small scales are sensitive to the last stages of inflation, where near-inflection trajectories and transient non-attractor evolution can amplify \(\zeta\) without changing the CMB band \citep{GarciaBellido:2017pbh,Germani:2017pbh}. Localized dips and controlled transitions provide concrete examples of such small-scale enhancement \citep{Briaud:2025transitions,Fujita:2025dip}. Related structures occur in monodromic valleys, dilaton-flattened potentials, and anomaly-inspired effective descriptions \citep{Pirzada:2026jml,Pirzada:2026sle,Pirzada:2026uak}. The open question is observational: which measured quantities can identify the same small-scale feature without relying on a primordial-black-hole tail?

Two low moments of \(\Pk\) give a sharp answer. Scalar inhomogeneities produce conformal fermions through a cubic functional of the curvature spectrum, so the relic abundance measures the integrated ultraviolet shoulder of the peak \citep{Garani:2025sdm,Garani:2025gmp}. The same perturbations source scalar-induced gravitational waves through second-order perturbation theory \citep{Acquaviva:2002ud,Mollerach:2003nq}. Radiation-era tensor kernels fix the quadratic convolution \citep{Ananda:2006af,Baumann:2007zm}. PBH and localized-spectrum studies supply the standard small-scale setting for this calculation \citep{Saito:2008jc,Kohri:2018awv,Pi:2020otn}. Reconstruction work has clarified how an observed stochastic spectrum constrains \(\Pk\) \citep{Domenech:2021ztg,Iovino:2025sigw,Ijaz:2024zma}. The new step here is to fix the scalar normalization with \(\Omega_\chi h^2\), then predict the MHz tensor amplitude from the same peak.

Eliminating the scalar area gives the relation
\begin{equation}
\begin{aligned}
 h^2\OmGW^{\rm pk}
 &= C_V\,
 \frac{(\Omchi h^2)^2}
 {(M_\chi/{\rm GeV})^2(\fzero/{\rm MHz})^6} \\
 &\quad\times
 \frac{1}{\Delta^2\exp(9\Delta^2)\Rthree^2},
\end{aligned}
 \label{eq:intro_relic_map}
\end{equation}
where \(\Delta\) is the logarithmic width of the scalar peak and \(\Rthree\) is the cubic-moment shape factor measured from the numerical spectrum. Equation~(\ref{eq:intro_relic_map}) is the central prediction. Once the relic abundance fixes the scalar area, the MHz signal is set by \(M_\chi\), \(f_0\), \(\Delta\), and \(\Rthree\). A null HFGW search gives a lower limit on \(M_\chi\) at fixed peak frequency. A detected stochastic background is overconstrained: the amplitude, peak location, spectral width, and relic abundance all have to arise from the same \(\Pk(k)\).

This mechanism belongs to the broader class of nonthermal dark-matter production. Superheavy gravitational relics established that the abundance can be set by cosmological dynamics instead of freeze-out \citep{Chung:1998ua,Chung:1998zb,Giudice:2000ex}. Freeze-in and Planck-suppressed production extend the same logic to very weakly coupled dark sectors \citep{Garny:2015sjg,Ema:2018ucl,Ema:2019yrd}. Quenched axion--\(SU(2)\) production and parametric-resonance QCD-axion production show how relic abundances can retain detailed information about time-dependent sources \citep{Redi:2021hsd,Redi:2022qrm,Khan:2026nsz}. Conformal production from scalar inhomogeneities is especially restrictive because the gravitational source that populates the fermions also generates a stochastic tensor background \citep{Pirzada:2026npl}.

High-frequency gravitational-wave searches now probe the relevant MHz band. Recent reviews summarize experimental concepts above \(10\,\mathrm{kHz}\) \citep{Aggarwal:2025hfgw}. We use the stochastic and power-law-integrated sensitivity conventions of HFGWplotter Omega and the strain compilation of HFGWplotter Sh \citep{Muia:2025Omega,Muia:2025Sh}. ABRACADABRA-10\,cm already covers the \(10\,\mathrm{kHz}\)--\(5\,\mathrm{MHz}\) interval containing the benchmark peak \citep{Pappas:2025abra}. Axion-haloscope and radio-conversion proposals provide complementary MHz response functions \citep{Domcke:2020radio,Domcke:2022haloscope,DMRadio:2022status,DMRadio:2025overview}. Broadband circuits, SRF cavities, and MAGO resonators extend the MHz comparison \citep{AlShirawi:2025dmr,Domcke:2025mwb,FLASH:2024cdr,Wenskat:2026srf}. Lunar concepts, BabyIAXO/RADES, and BREAD connect the same frequency range to larger electromagnetic instruments \citep{Jani:2020gloc,Valero:2025babyiaxo,Capdevilla:2025bread}. Multimode cavities and detector networks add spectral and geographical redundancy \citep{Blas:2026multimode,Amaral:2026gravnet}. Spin, qumode, and first-order-transition searches extend the comparison to distinct microscopic couplings \citep{Liang:2025spin,Kharzeev:2025qugrav,Ai:2025gtr}. Our fiducial convolution peaks at \(2.63\,\mathrm{MHz}\); the strongest nearby deformation peaks at \(2.04\,\mathrm{MHz}\). Both lie inside the laboratory band already being mapped by current and proposed searches.

The rest of the paper develops the calculation. Section~\ref{sec:moment-map} defines the moments of \(\Pk\), derives the cubic conformal-fermion kernel, and fixes the relic normalization. Section~\ref{sec:sigw} evaluates the radiation-era SIGW convolution. Section~\ref{sec:closure} derives the abundance-normalized closure relation and mass--frequency inference. Section~\ref{sec:hfgw-translation} translates the result into HFGW observables. Section~\ref{sec:inflation} supplies a Mukhanov--Sasaki benchmark, Sec.~\ref{sec:control} checks reheating, PBH, dark-radiation, and dark-sector consistency, and Sec.~\ref{sec:robustness} studies controlled deformations of the peak.

\begin{table*}[tbp]
\centering
\small
\begin{tabular}{@{}p{0.20\textwidth}p{0.25\textwidth}p{0.25\textwidth}p{0.18\textwidth}@{}}
\toprule
Line of work & Established result & Limitation for this problem & Role in this work \\
\midrule
PBH constraints & Enhanced small-scale power can collapse at horizon entry. & Tail, threshold, and profile uncertainties obscure smooth sub-PBH peaks. & PBH constraints provide a consistency check instead of the defining observable. \\
Independently normalized SIGW templates & A scalar peak induces a stochastic tensor background. & The scalar amplitude is usually fitted or scanned independently. & The amplitude is fixed by the dark-matter abundance. \\
Conformal production from inhomogeneities & Curvature perturbations produce conformal fermions through a cubic moment. & Tensor normalization was independent in those calculations. & The cubic yield is closed with the SIGW kernel. \\
HFGW detector coverage & MHz--GHz searches now have published sensitivity conventions. & Sensitivity curves require a source-specific cosmological signal. & The closure relation translates them into mass reach. \\
\bottomrule
\end{tabular}
\caption{Relation to the surrounding literature. The kernels are standard; the new ingredient is the relic-normalized closure between the cubic dark-matter moment and the quadratic tensor moment.}
\label{tab:literature-position}
\end{table*}

\begin{figure*}[tbp]
\centering
\includegraphics[width=0.98\textwidth]{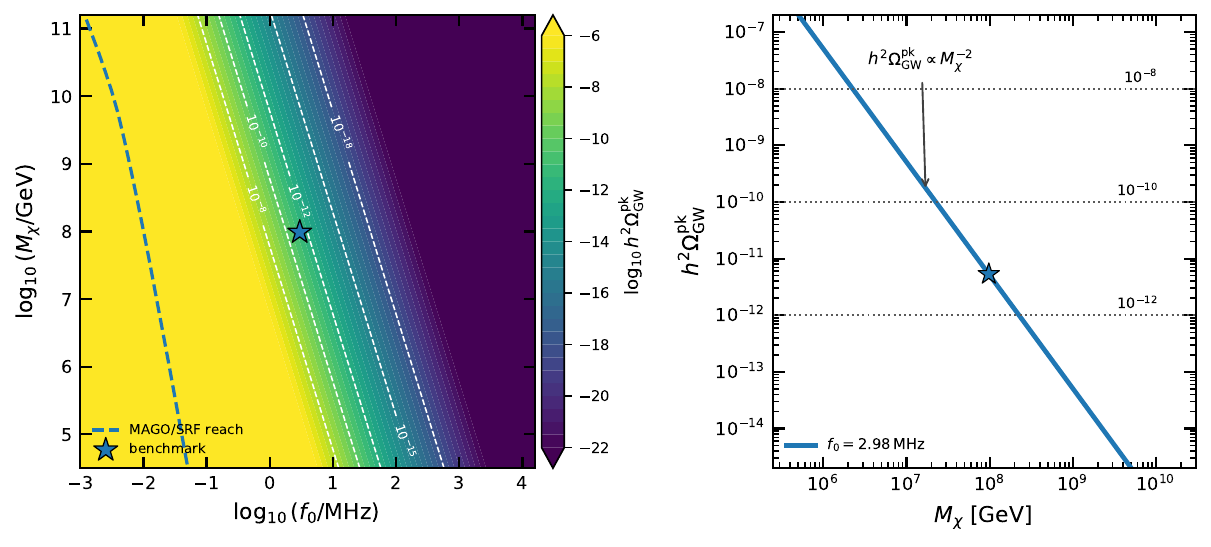}
\caption{Relic-normalized MHz signature. Left: predicted tensor peak after imposing \(\Omega_\chi h^2=0.12\), shown in the \((f_0,M_\chi)\) plane for the benchmark \(\Delta\), \(\Rthree\), and \(C_V\). White contours give fixed \(h^2\Omega_{\rm GW}^{\rm pk}\); dashed curves translate representative PLI sensitivities into mass reach. Right: at the benchmark frequency the signal follows the steep \(M_\chi^{-2}\) law. The star marks the direct Mukhanov--Sasaki/SIGW benchmark.}
\label{fig:signature-map}
\end{figure*}

\section{Primordial moment map}
\label{sec:moment-map}

\subsection{Spectrum and moments}

We start from the curvature spectrum at the end of inflation. The numerical input is \(\Pk^{\rm num}(k)\), obtained by solving the Mukhanov--Sasaki equation in Sec.~\ref{sec:inflation}. For analytic bookkeeping we compress the localized peak into the lognormal profile
\begin{equation}
\begin{aligned}
 \Pk^{\rm LN}(k)
 &=
 \frac{\Az}{\sqrt{2\pi}\Delta}
 \exp\left[-\frac{\ln^2(k/k_0)}{2\Delta^2}\right],\\
 \int \dd\ln k\,\Pk^{\rm LN}(k)&=\Az .
\end{aligned}
 \label{eq:lognormal}
\end{equation}
Here \(\Az\) is the integrated peak area, \(k_0\) the characteristic scale, and \(\Delta\) the logarithmic width. The lognormal form is only a dictionary for the parameter dependence. The relic abundance and the induced tensor spectrum are computed from \(\Pk^{\rm num}(k)\).

The present-day frequency corresponding to a comoving scale is
\begin{equation}
 f[\mathrm{MHz}]
 =
 1.546\times10^{-21}\,
 k[\kMpc],
 \label{eq:freq}
\end{equation}
so \(k_0=1.927\times10^{21}\,\kMpc\) gives \(\fzero=2.979\,\mathrm{MHz}\).

The lognormal moments are
\begin{equation}
 \mathcal{M}_n^{\rm LN}
 \equiv
 \int\dd\ln k\,k^n\Pk^{\rm LN}(k)
 =
 \Az k_0^n
 \exp\left(\frac{n^2\Delta^2}{2}\right),
 \label{eq:mn}
\end{equation}
so each observable selects a different weighted part of the peak. For the cubic moment relevant to conformal-fermion production, setting \(y=\ln(k/k_0)\) gives
\begin{equation}
 3y-\frac{y^2}{2\Delta^2}
 =
 -\frac{(y-3\Delta^2)^2}{2\Delta^2}
 +\frac{9\Delta^2}{2},
 \label{eq:saddle}
\end{equation}
with the saddle on the ultraviolet shoulder. A peak with \(\Pk^{\rm max}\sim10^{-3}\) can then have a small Gaussian PBH tail and still carry a large cubic moment.

For the numerical spectrum we keep the exact cubic moment,
\begin{equation}
 \Mthree^{\rm num}
 \equiv
 \int\dd\ln k\,k^3\Pk^{\rm num}(k),
 \quad
 \Rthree
 \equiv
 \frac{\Mthree^{\rm num}}
 {\Az k_0^3
 \exp(9\Delta^2/2)} .
 \label{eq:Rthree}
\end{equation}
The benchmark gives \(\Rthree=1.6287\). This number measures how far the Mukhanov--Sasaki spectrum departs from the moment-matched lognormal profile; recomputing it for every numerical solution keeps the shape dependence explicit.

The normalization, width, and shape dependence separate cleanly by writing
\begin{equation}
 \Pk(k)
 =
 \frac{\Az}{\Delta}
 S\!\left(
 \frac{\ln(k/k_0)}{\Delta}
 \right),
 \qquad
 \int_{-\infty}^{\infty}
 \dd x\,S(x)=1 .
 \label{eq:shape-decomposition}
\end{equation}
For an ideal lognormal spectrum,
\(
S(x)
=
(2\pi)^{-1/2}
\exp(-x^2/2),
\)
whereas the numerical calculation uses the normalized Mukhanov--Sasaki profile. The conformal-fermion abundance scales as \(\Omega_\chi h^2\propto\Az\), while the SIGW amplitude scales as \(h^2\Omega_{\rm GW}\propto\Az^2\). Fixing \(\Az\) with the relic density removes the free scalar normalization and leaves the tensor signal controlled by \(\Delta\), \(\Rthree\), and the quadratic tensor shape factor.

\subsection{Fermion production and normalization}

Conformal fermions are produced only by inhomogeneities. In a homogeneous FLRW spacetime, the Weyl rescaling \(\psi=a^{3/2}\chi\) reduces the massless fermion action to the Minkowski form, so the cosmological expansion alone creates no particles. Scalar perturbations break that cancellation through spatial gradients and constraint fields in the perturbed geometry.

Working in comoving gauge, the perturbed metric is
\begin{equation}
 \dd s^2
 =
 a^2(\tau)
 \left[
 -(1+2\Phi)\dd\tau^2
 +
 \ee^{2\zeta(\tau,\mathbf{x})}
 \dd\mathbf{x}^2
 \right],
 \label{eq:scalar-metric}
\end{equation}
and the interaction Hamiltonian of the rescaled fermion contains
\begin{equation}
\begin{aligned}
 H_{\rm int}(\tau)
 &=
 \int\dd^3x\,\bar\psi
 \Big[
 c_1\zeta'\gamma^0
 +c_2\partial_i\zeta\gamma^i\\
 &\qquad
 +c_3\partial_i\Phi\gamma^0\gamma^i
 +\cdots
 \Big]\psi .
\end{aligned}
 \label{eq:fermion-hint}
\end{equation}
The leading pair-production amplitude is the Bogoliubov coefficient
\begin{equation}
\begin{aligned}
 \beta_{\mathbf{p}\mathbf{q}}
 &=
 -i
 \int\dd\tau\,
 \langle
 \mathbf{p},\mathbf{q}
 |
 H_{\rm int}(\tau)
 |0\rangle,\\
 n_\chi a^3
 &=
 2
 \int
 \frac{\dd^3p}{(2\pi)^3}
 \frac{\dd^3q}{(2\pi)^3}
 |\beta_{\mathbf{p}\mathbf{q}}|^2 .
\end{aligned}
 \label{eq:beta-fermion}
\end{equation}
This expression counts the fermion pairs created by the time-dependent inhomogeneous metric.

Solving the scalar constraints and integrating over fermion phase space reduces the production rate to a local functional of \(\Pk\). Rotational invariance supplies the measure \(k^3\dd\ln k\), and conformal invariance removes explicit factors of the scale factor, Hubble rate, and fermion mass during production. The kernel is \citep{Garani:2025sdm,Garani:2025gmp}
\begin{equation}
 n_\chi a^3
 =
 \frac{\Achi}{4\pi^2}
 \int\dd\ln k\,k^3\Pk(k),
 \qquad
 \Achi\simeq0.015,
 \label{eq:nchi}
\end{equation}
where \(\Achi\) contains the spin sum, time integral, and constraint normalization. The cubic dependence follows from the production amplitude and phase-space geometry. It also explains why the fermion abundance samples the ultraviolet-weighted body of the peak, while the tensor channel probes a quadratic convolution.

With \(N_\chi\equiv n_\chi a^3\), the late mass-generating transition gives \(\rho_{\chi,0}=M_\chi N_\chi/a_0^3\). Combining Eq.~\eqref{eq:nchi}, Eq.~\eqref{eq:Rthree}, and Eq.~\eqref{eq:freq} yields
\begin{equation}
 \Omchi h^2
 =
 C_\chi
 \left(
 \frac{M_\chi}{\mathrm{GeV}}
 \right)
 \left(
 \frac{\fzero}{\mathrm{MHz}}
 \right)^3
 \Az
 \exp\left(\frac{9\Delta^2}{2}\right)
 \Rthree ,
 \label{eq:omchi}
\end{equation}
which relates the relic density to the integrated peak area and the cubic shape factor.

The normalization constant is
\begin{equation}
 C_\chi
 =
 \frac{\Achi}{4\pi^2}
 \frac{(1.546\times10^{-21})^{-3}}
 {(3.0856776\times10^{24})^3}
 \frac{1}{\rho_c/h^2}
 =
 3.3206\times10^{-10},
 \label{eq:cchi}
\end{equation}
with
\(
\rho_c/h^2
 =
1.05375\times10^{-5}\,
\mathrm{GeV\,cm^{-3}}
\).
The factors convert the frequency to \(\mathrm{Mpc}^{-1}\), the comoving number density to \(\mathrm{cm}^{-3}\), and the resulting mass density to \(\rho_c/h^2\).

The relic abundance fixes the integrated scalar area,
\begin{equation}
 \Az
 =
 \frac{\Omchi h^2}
 {C_\chi
 (M_\chi/\mathrm{GeV})
 (\fzero/\mathrm{MHz})^3
 \exp(9\Delta^2/2)
 \Rthree},
 \label{eq:area-from-dm}
\end{equation}
for a chosen fermion mass and measured peak shape. This is the normalization step that converts the SIGW calculation into a prediction.

\begin{figure*}[tbp]
\centering
\includegraphics[width=0.95\textwidth]{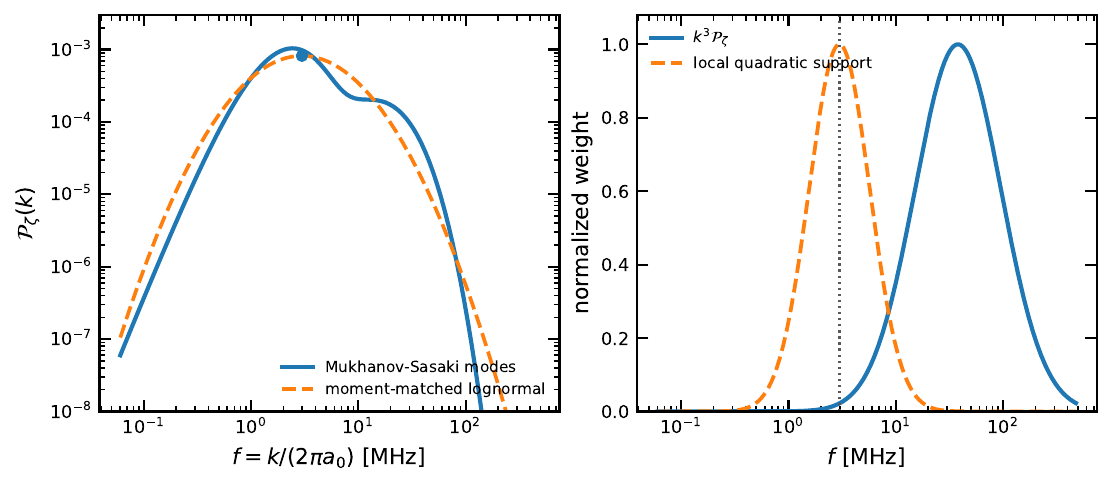}
\caption{Scalar spectrum and moment weights. Left: Mukhanov-Sasaki curvature spectrum around the amplified band and its moment-matched lognormal compression. Right: normalized support of the cubic weight \(k^3\Pk\) compared with a local quadratic support proxy. The fermion channel is displaced toward the ultraviolet shoulder, while the tensor channel is controlled by a quadratic convolution of modes re-entering together.}
\label{fig:spectrum-moments}
\end{figure*}

\begin{table}[tbp]
\centering
\caption{Tensor-kernel convergence for the benchmark. The parameter \(L\) is the half-width of the logarithmic integration domain and \(N\) denotes the quadrature resolution in each logarithmic direction.}
\label{tab:convergence}
\small
\begin{tabular}{cccc}
\toprule
\(N\) & \(L\) & \(h^2\Omega_{\rm GW}^{\rm pk}\) & Relative shift \\
\midrule
220 & 5.5 & \(5.132\times10^{-12}\) & \(3.77\%\) \\
260 & 6.0 & \(4.965\times10^{-12}\) & \(6.89\%\) \\
300 & 6.0 & \(5.242\times10^{-12}\) & \(1.70\%\) \\
340 & 6.0 & \(5.333\times10^{-12}\) & reference \\
300 & 6.5 & \(5.216\times10^{-12}\) & \(2.19\%\) \\
\bottomrule
\end{tabular}
\end{table}

\section{Induced gravitational waves}
\label{sec:sigw}

The same \(\Pk(k)\) sources a stochastic gravitational-wave background at second order. Standard SIGW calculations often scan the scalar amplitude as a free normalization. Here the scalar area has already been fixed by Eq.~\eqref{eq:area-from-dm}, so the tensor spectrum follows from the radiation-era convolution applied to the relic-normalized peak. We use the standard second-order treatment for radiation domination \citep{Acquaviva:2002ud,Mollerach:2003nq,Ananda:2006af,Baumann:2007zm} and the localized-spectrum conventions developed in PBH and reconstruction studies \citep{Saito:2008jc,Kohri:2018awv,Pi:2020otn,Domenech:2021ztg}.

During radiation domination, the present-day gravitational-wave energy density is given by
\begin{widetext}
\begin{equation}
h^2\OmGW(k)
=
\frac{\mathcal{T}_g h^2\Omega_{r,0}}{24}
\int_0^\infty \dd v
\int_{|1-v|}^{1+v}\dd u\,
\mathcal{K}(u,v)\,
\overline{I^2(u,v)}
\Pk(ku)\Pk(kv),
\label{eq:sigw-master}
\end{equation}
\end{widetext}
where \(h^2\Omega_{r,0}=4.18\times10^{-5}\) and \(\mathcal{T}_g\simeq0.83\) accounts for the evolution of the relativistic degrees of freedom between production and the present epoch. The geometric mode-coupling kernel is
\begin{equation}
\mathcal{K}(u,v)
=
\left[
\frac{
4v^2-(1+v^2-u^2)^2
}
{4uv}
\right]^2 ,
\label{eq:geomkernel}
\end{equation}
while the corresponding time-averaged radiation-era transfer function is
\begin{widetext}
\begin{align}
\overline{I^2(u,v)}
&=
\frac12
\left[
\frac{
3(u^2+v^2-3)
}
{4u^3v^3}
\right]^2
\Bigg\{
\left[
-4uv
+
(u^2+v^2-3)
\ln
\left|
\frac{3-(u+v)^2}
{3-(u-v)^2}
\right|
\right]^2
\nonumber\\
&\hspace{4cm}
+\pi^2
(u^2+v^2-3)^2
\Theta(u+v-\sqrt3)
\Bigg\}.
\label{eq:ibarsq}
\end{align}
\end{widetext}

Equations~\eqref{eq:sigw-master}--\eqref{eq:ibarsq} fix the induced tensor spectrum for any specified \(\Pk\). We evaluate the convolution directly with \(\Pk^{\rm num}\). Varying the logarithmic integration range and the \((u,v)\) quadrature gives the convergence shown in Table~\ref{tab:convergence}: the peak amplitude shifts by less than \(6.9\%\) across the full grid set, and the two finest grids at fixed range differ by \(1.7\%\).

The broad logarithmic width of the scalar peak produces a broad SIGW spectrum. Extracting the maximum directly from the numerical convolution gives
\begin{equation}
f_{\rm GW}^{\rm pk}
=
2.6275\,\mathrm{MHz},
\qquad
h^2\OmGW^{\rm pk}
=
5.3325\times10^{-12},
\label{eq:gw-benchmark}
\end{equation}
with
\(
\xi\equiv
f_{\rm GW}^{\rm pk}/\fzero
=
0.88197.
\)

Using Eq.~\eqref{eq:shape-decomposition}, the peak tensor amplitude has the form
\begin{equation}
h^2\OmGW^{\rm pk}
=
\frac{\Az^2}{\Delta^2}
\,
\mathcal{Q}_2[S;\Delta,\xi],
\label{eq:qtwo}
\end{equation}
where \(\mathcal{Q}_2\) is the radiation-era double convolution evaluated at the tensor maximum. The \(\Az^2\) scaling follows from the bilinear source term in Eq.~\eqref{eq:sigw-master}; the conformal-fermion abundance in Eq.~\eqref{eq:omchi} scales linearly with \(\Az\).

For the benchmark spectrum,
\begin{equation}
h^2\OmGW^{\rm pk}
=
\mathcal{C}_{\rm GW}
\frac{\Az^2}{\Delta^2},
\qquad
\mathcal{C}_{\rm GW}
=
1.2665\times10^{-6},
\label{eq:cgw}
\end{equation}
where \(\mathcal{C}_{\rm GW}\) is the quadratic shape functional of the numerical peak. It is the tensor analogue of \(\Rthree\): \(\Rthree\) weights the cubic fermion channel, while \(\mathcal{C}_{\rm GW}\) weights the quadratic tensor channel. Keeping both quantities explicit prevents the peak morphology from being absorbed into an effective amplitude or mass. This forward convolution complements reconstruction and spectral-sharpness analyses of induced gravitational waves \citep{ElGammal:2025sigway,Inomata:2024sharp}.

\begin{widetext}
To facilitate applications beyond the benchmark model, the quadratic functional may be parameterized analytically for ideal lognormal spectra. Evaluating the radiation-era kernel over the interval \(0.45\leq\Delta\leq1.35\) yields
\begin{equation}
\mathcal{C}_{\rm GW}^{\rm LN}(\Delta)
\simeq
8.7556\times10^{-7}
\exp\!\left[
0.396(\Delta-\Delta_{\rm fid})
-
0.394(\Delta-\Delta_{\rm fid})^2
\right],
\qquad
\Delta_{\rm fid}=0.92096,
\label{eq:cgw-lognormal-fit}
\end{equation}
whose maximum fractional deviation from the numerical scan is only \(1.6\%\). The benchmark spectrum differs from an exact lognormal, so its quadratic functional is written as
\begin{equation}
\mathcal{C}_{\rm GW}^{\rm num}(\Delta,S)
=
\mathcal{S}_2[S]
\,
\mathcal{C}_{\rm GW}^{\rm LN}(\Delta),
\qquad
\mathcal{S}_2[S_{\rm fid}]
=
\frac{1.2665\times10^{-6}}
{8.7556\times10^{-7}}
=
1.446,
\label{eq:cgw-shape2}
\end{equation}
where the dimensionless factor \(\mathcal{S}_2\) quantifies the departure of the numerical spectrum from the ideal lognormal shape. Nearby primordial spectra may be treated through a direct numerical convolution, as done throughout the benchmark analysis, or through the analytical approximation in Eq.~\eqref{eq:cgw-lognormal-fit} supplemented by the measured shape correction \(\mathcal{S}_2\). This decomposition separates the universal dependence on the peak width from the genuine spectral-shape information, making the quadratic functional readily transferable to other localized primordial spectra without sacrificing the predictive structure of the abundance-normalized calculation.
\end{widetext}

\section{Abundance-normalized closure}
\label{sec:closure}

Eliminating \(\Az\) between Eq.~\eqref{eq:omchi} and Eq.~\eqref{eq:cgw} gives
\begin{equation}
\begin{aligned}
 h^2\OmGW^{\rm pk}
 &=
 C_V
 \frac{(\Omchi h^2)^2}
 {(M_\chi/\mathrm{GeV})^2(\fzero/\mathrm{MHz})^6}
 \\
 &\quad\times
 \frac{1}
 {\Delta^2
 \exp(9\Delta^2)
 \Rthree^2},
 \qquad
 C_V\equiv
 \frac{\mathcal{C}_{\rm GW}}{C_\chi^2},
\end{aligned}
\label{eq:relicmap}
\end{equation}
with \(C_V=1.1486\times10^{13}\) for the benchmark. The relation fixes the SIGW peak after the relic abundance is imposed. The main scaling,
\(h^2\Omega_{\rm GW}^{\rm pk}\propto M_\chi^{-2}f_0^{-6}\),
gives strong leverage to MHz searches, while the factor \(\exp(9\Delta^2)\) penalizes broad scalar peaks.

The logarithmic response makes the parameter dependence explicit,
\begin{widetext}
\begin{align}
\dd\ln(h^2\OmGW^{\rm pk})
&=
-2\,\dd\ln M_\chi
-6\,\dd\ln\fzero
-2\,\dd\ln\Rthree
-(2+18\Delta^2)\dd\ln\Delta
\nonumber\\
&\qquad
+\dd\ln C_V
+2\,\dd\ln(\Omchi h^2),
\label{eq:logresponse}
\end{align}
\end{widetext}
so the fiducial \(\Delta=0.92096\) gives a width coefficient \(-17.3\,\dd\ln\Delta\). A measured stochastic spectrum would constrain the mass and frequency scales, but its width would carry comparable information through \(\Delta\) and \(\Rthree\).

For an upper limit or projected sensitivity \(\Omega_{\rm lim}(f)\) evaluated at \(f=f_{\rm GW}^{\rm pk}\simeq\xi\fzero\), Eq.~\eqref{eq:relicmap} gives the mass reach
\begin{equation}
\begin{aligned}
M_\chi^{\rm lim}(\fzero)
&=
\left[
\frac{
C_V(\Omchi h^2)^2
}
{
\Omega_{\rm lim}(\xi\fzero)
(\fzero/\mathrm{MHz})^6
}
\right]^{1/2}
\\
&\quad\times
\frac{\mathrm{GeV}}
{\Delta
\exp(9\Delta^2/2)
\Rthree},
\end{aligned}
\label{eq:massreach}
\end{equation}
where the inverse quadratic dependence on \(M_\chi\) converts stronger HFGW limits into stronger lower bounds on the fermion mass. The translation uses the measured relic abundance and introduces no second scalar-amplitude parameter.

A stochastic detection can be inverted to an inferred dark-matter mass,
\begin{equation}
\begin{aligned}
M_\chi^{\rm inf}
&=
\frac{\Omchi h^2}
{\Delta
\exp(9\Delta^2/2)
\Rthree}
\left[
\frac{f_{\rm GW}^{\rm pk}}{\xi\,\mathrm{MHz}}
\right]^{-3}\\
&\quad\times
\left(
\frac{C_V}
{h^2\Omega_{\rm GW}^{\rm pk}}
\right)^{1/2}
\mathrm{GeV} .
\end{aligned}
\label{eq:mass-inference}
\end{equation}
so one mass is required to reproduce the relic abundance and the tensor peak for one primordial spectrum.

The residual shape dependence follows from

\begin{equation}
\begin{aligned}
\frac{
M_\chi(\Delta,\Rthree,\mathcal{S}_2)
}
{
M_\chi(\Delta_{\rm fid},
\mathcal{R}_{3,\rm fid},
\mathcal{S}_{2,\rm fid})
}
=\\
\left[
\frac{
\mathcal{C}_{\rm GW}
(\Delta,\mathcal{S}_2)
}
{
\mathcal{C}_{\rm GW}
(\Delta_{\rm fid},
\mathcal{S}_{2,\rm fid})
}
\right]^{1/2}
\frac{
\Delta_{\rm fid}
\mathcal{R}_{3,\rm fid}
\exp(9\Delta_{\rm fid}^2/2)
}
{
\Delta
\Rthree
\exp(9\Delta^2/2)
},
\end{aligned}
\label{eq:shape-mass-ratio}
\end{equation}

which separates the quadratic tensor factor from the cubic fermion factor. Figure~\ref{fig:closure-diagnostic} shows the corresponding inversion in the observed \((f_{\rm GW}^{\rm pk},h^2\Omega_{\rm GW}^{\rm pk})\) plane and the mass shift induced by changing \(\Delta\) and \(\Rthree\). A candidate MHz signal passes the closure test only if a common \(\Pk(k)\) accounts for the relic density, tensor amplitude, peak frequency, and width.

\begin{figure*}[tbp]
\centering
\includegraphics[width=0.9\textwidth]{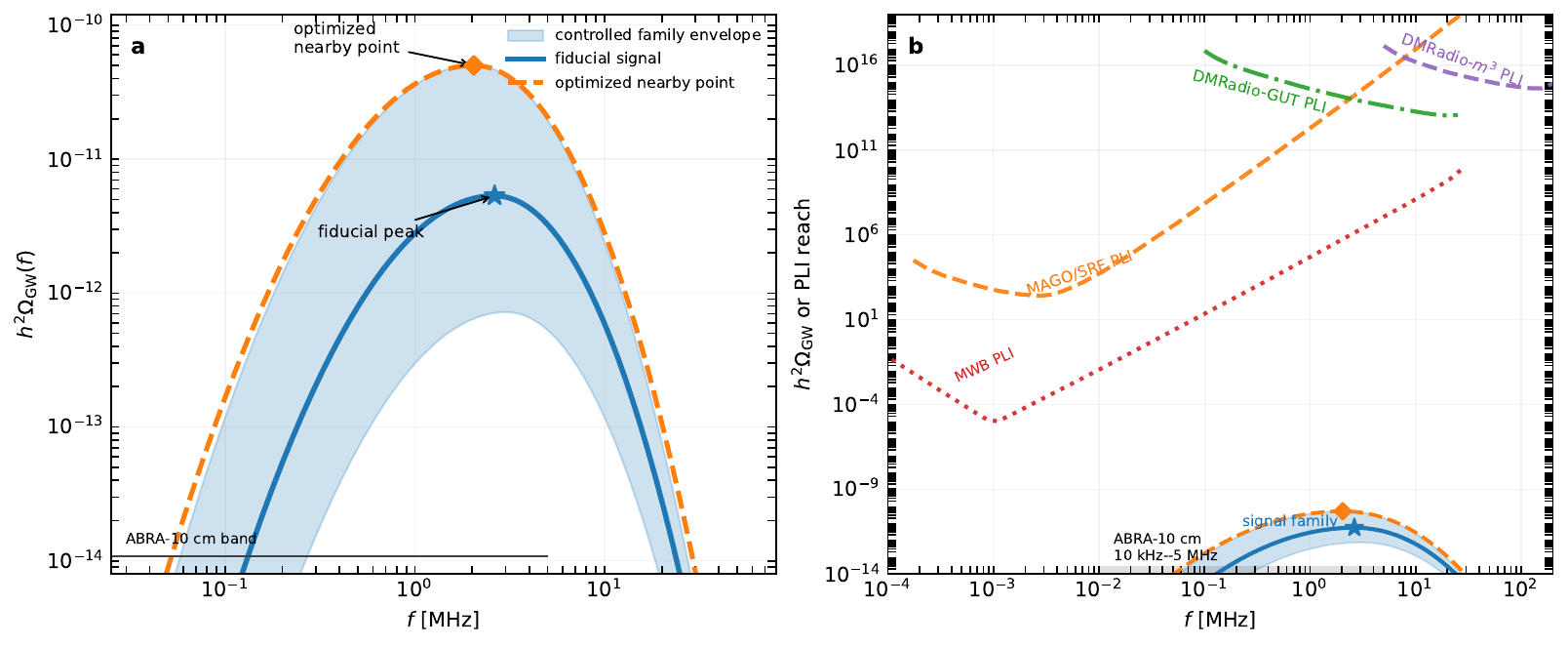}
\caption{Induced gravitational-wave spectrum and MHz sensitivity comparison. Left: direct radiation-era convolution for the fiducial point and the bounded envelope of the nearby family \(D=17.2\)--\(18.8\); the star marks the fiducial peak and the diamond marks the largest-amplitude nearby point. Right: representative stochastic/PLI sensitivities overlaid on the same family. The fill covers only the region between the family curves, and the ABRACADABRA interval is a compact published search band. PLI curves indicate reach; likelihood exclusions depend on the experiment-specific stochastic estimator and response convention.}
\label{fig:gw-sensitivity}
\end{figure*}

\begin{figure*}[tbp]
\centering
\includegraphics[width=14cm, height=8cm]{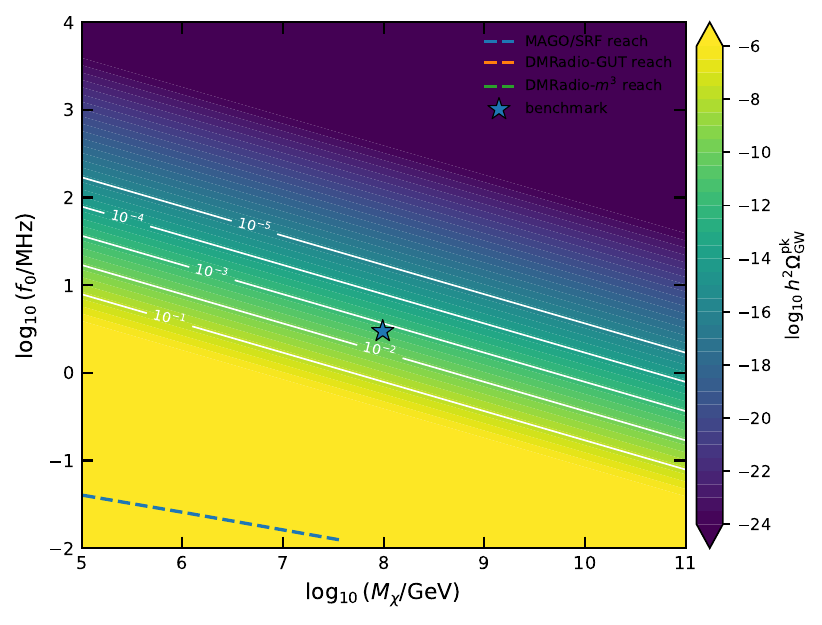}
\caption{Relic-normalized prediction in the \((M_\chi,\fzero)\) plane for the benchmark \(\Delta\), \(\Rthree\), and \(C_V\). The color gives the peak tensor amplitude after imposing \(\Omchi h^2=0.12\). White contours show the scalar area required by the relic abundance. Dashed curves translate representative PLI sensitivities into mass reach.}
\label{fig:relic-map}
\end{figure*}

\begin{table}[tbp]
\centering
\caption{High-frequency curve conventions used in the figures. Current limits and projections are separated; PLI curves are reach indicators unless an experiment supplies a likelihood or an upper-limit curve for the benchmark spectral shape.}
\label{tab:curves}
\tiny
\setlength{\tabcolsep}{1.2pt}
\begin{tabular}{@{}p{0.20\columnwidth}p{0.25\columnwidth}p{0.12\columnwidth}p{0.30\columnwidth}@{}}
\toprule
Curve family & Native convention & Status & Use in this paper \\
\midrule
MAGO/SRF PLI & stoch. \(h^2\Omega_{\rm GW}\) reach & Proj. & MHz reach and mass translation \\
DMRadio-GUT PLI & stoch. \(h^2\Omega_{\rm GW}\) reach & Proj. & high-frequency reach \\
DMRadio-\(m^3\) PLI & stoch. \(h^2\Omega_{\rm GW}\) reach & Proj. & higher-MHz scan \\
IAXO-LF/IAXO PLI & stoch. \(h^2\Omega_{\rm GW}\) reach & Proj. & wide-frequency program \\
ABRA-10 cm & published MHz search band & Current & benchmark-band context \\
\bottomrule
\end{tabular}
\end{table}

\section{Experimental translation}
\label{sec:hfgw-translation}

The closure relation predicts \(h^2\Omega_{\rm GW}(f)\) in terms of the curvature peak and the dark sector. Comparing it with laboratory searches requires the same observable convention used by the experiment. Figures~\ref{fig:gw-sensitivity} and~\ref{fig:relic-map} use the curve classes summarized in Table~\ref{tab:curves}. We take stochastic and power-law-integrated sensitivities from HFGWplotter Omega, and strain curves from HFGWplotter Sh \citep{Muia:2025Omega,Muia:2025Sh}. The conversion follows the standard stochastic-background formalism \citep{Thrane:2013oya,Tamarit:2025hfgwplotter}. Recent HFGW reviews give the wider experimental context above \(10\,\mathrm{kHz}\) \citep{Aggarwal:2025hfgw}.

The quantities \(h^2\Omega_{\rm GW}\), \(h_c\), and \(S_h\) refer to different stochastic observables. For an isotropic background,
\begin{equation}
 \Omega_{\rm GW}(f)
 =
 \frac{2\pi^2}{3H_0^2}
 f^3
 S_h(f),
 \qquad
 h_c^2(f)
 =
 fS_h(f),
 \label{eq:strainomega}
\end{equation}
up to the response normalization. Curves already reported as stochastic or PLI \(h^2\Omega_{\rm GW}\) can be compared directly with the prediction. Strain curves require Eq.~\eqref{eq:strainomega}, the observation time, the detector response, and the stochastic estimator convention.

Different HFGW detector concepts probe the same stochastic background with different response functions and noise systematics. For a detector pair, the standard cross-correlation statistic is
\begin{widetext}
\begin{equation}
 \hat Y_{ij}
 =
 \int\dd f\,
 \tilde s_i^*(f)
 Q_{ij}(f)
 \tilde s_j(f),
 \qquad
 {\rm SNR}_{ij}^2
 \simeq
 2T
 \int\dd f\,
 \frac{
 \Gamma_{ij}^2(f)
 \Omega_{\rm GW}^2(f)
 }
 {
 f^6
 P_i(f)
 P_j(f)
 }
 \left(
 \frac{3H_0^2}{10\pi^2}
 \right)^2 ,
 \label{eq:crosscorr}
\end{equation}
\end{widetext}
where \(P_i(f)\) and \(P_j(f)\) are the noise power spectra, \(Q_{ij}(f)\) is the optimal filter, and \(\Gamma_{ij}(f)\) is the overlap reduction function. Resonant cavities and broadband circuits have natural MHz sensitivity to narrow stochastic features \citep{Domcke:2020radio,Domcke:2022haloscope,Wenskat:2026srf}. Magnon, spin, and qumode devices test the same background with different instrumental couplings \citep{Liang:2025spin,Kharzeev:2025qugrav}. A cross-platform measurement of the width would constrain \(\Delta\), \(\Rthree\), and \(\mathcal{C}_{\rm GW}\), reducing the shape degeneracy in Fig.~\ref{fig:closure-diagnostic}.

Equation~\eqref{eq:massreach} is the experimental translation of the closure relation. Current limits and projected sensitivities define regions in the \((M_\chi,\fzero)\) plane after \(\Omega_\chi h^2\) fixes the scalar normalization. A stochastic detection would determine more than a tensor amplitude: the measured frequency, width, and intensity would jointly constrain the primordial scale, the inferred fermion mass, and the peak geometry. The sensitivity curves in the figures are shown for that consistency test.

\begin{table}[tbp]
\centering
\caption{Benchmark values obtained from the direct numerical spectrum and radiation-era tensor convolution.}
\label{tab:benchmark}
\scriptsize
\setlength{\tabcolsep}{3pt}
\begin{tabular}{lc}
\toprule
Quantity & Value \\
\midrule
Scalar area \(\Az\) & \(1.8898\times10^{-3}\) \\
Width \(\Delta\) & \(0.92096\) \\
Direct cubic shape factor \(\Rthree\) & \(1.6287\) \\
Scalar central frequency \(\fzero\) & \(2.9791\,\mathrm{MHz}\) \\
GW peak frequency \(f_{\rm GW}^{\rm pk}\) & \(2.6275\,\mathrm{MHz}\) \\
Fermion mass from \(\Omchi h^2=0.12\) & \(9.77\times10^7\,\mathrm{GeV}\) \\
Peak tensor amplitude & \(5.33\times10^{-12}\) \\
Numerical peak height \(\Pk^{\rm max}\) & \(1.04\times10^{-3}\) \\
Integrated GW dark-radiation contribution & \(\Delta\Neff^{\rm GW}\ll10^{-10}\) \\
\bottomrule
\end{tabular}
\end{table}

\section{Single-field realization}
\label{sec:inflation}

The closure relation is independent of a unique inflationary microphysics. For the benchmark spectrum we use a local single-field background that generates a smooth small-scale enhancement while keeping the CMB band unchanged. The ansatz provides a controlled single-clock source for the Mukhanov--Sasaki spectrum and the SIGW convolution.

The first slow-roll parameter evolves as
\begin{equation}
 \ln\epsilon(N)=\ln\epsilon_\star+\beta N+\gamma N^2
 -D\exp\left[-\frac{(N-N_{\rm pk})^2}{2\sigma^2}\right],
 \label{eq:epsansatz}
\end{equation}
where \(N=0\) at the CMB pivot and \(N_{\rm end}=60\). The polynomial part fixes the large-scale slow-roll evolution; the localized Gaussian suppression reduces \(\epsilon\) and enhances scalar fluctuations over a finite band. The fiducial values are
\begin{widetext}
\begin{equation}
\epsilon_\star=2.2\times10^{-4},\quad
 \beta=0.0324,
 \quad \gamma=1.7994\times10^{-3},
 \quad D=18.0,
 \quad \sigma=1.4,
 \quad N_{\rm pk}=53.2 ,
 \label{eq:params}
\end{equation}
\end{widetext}
and the strongest nearby realization uses \(D=18.8\), giving \(h^2\Omega_{\rm GW}^{\rm pk}=5.01\times10^{-11}\) at \(2.04\,\mathrm{MHz}\).

Hamilton--Jacobi reconstruction gives \citep{Lidsey:1995np}
\begin{equation}
\begin{aligned}
 \frac{\dd\ln H}{\dd N}&=-\epsilon,
 &
 \frac{\dd\phi}{\dd N}&=\sqrt{2\epsilon}\,\Mpl,\\
 V(\phi)&=H^2\Mpl^2(3-\epsilon).
\end{aligned}
 \label{eq:hj}
\end{equation}
which fixes the local background over the feature. Scalar perturbations obey the Mukhanov--Sasaki equation \citep{Mukhanov:1990me,Stewart:1993bc},
\begin{equation}
 u_k''+\left(k^2-\frac{z''}{z}\right)u_k=0,
 \qquad
 z=a\sqrt{2\epsilon}\,\Mpl,
 \label{eq:ms}
\end{equation}
with Bunch--Davies initial data deep inside the horizon. After freeze-out, the same numerical \(\Pk(k)\) enters Eq.~\eqref{eq:Rthree} and Eq.~\eqref{eq:sigw-master}; the calculation introduces no independent tensor template or scalar renormalization.

The parameters \(D\), \(\sigma\), and \(N_{\rm pk}\) control different parts of the peak. \(D\) changes the depth of the transient slow-roll suppression and mainly changes \(\Az\). \(\sigma\) sets the logarithmic width. \(N_{\rm pk}\) fixes the comoving scale. These quantities describe the local feature; a single global potential may impose additional relations among them. Any smooth spectrum can be passed through the same closure calculation after recomputing \(\Delta\), \(\Rthree\), and \(\mathcal{C}_{\rm GW}\). The vacuum tensor spectrum follows the Hubble scale and remains subdominant on the MHz scales considered here; the enhanced signal comes from the scalar-induced channel.

The deformation parameters leave distinct signatures. Increasing \(D\) enlarges \(\Az\) and raises the induced signal before the abundance normalization is imposed. Changing \(\sigma\) alters \(\Delta\), giving
\begin{equation}
\begin{aligned}
 \frac{\dd\ln M_\chi}{\dd\ln\Delta}
 &=
 -1-9\Delta^2
 +\frac12\frac{\dd\ln\mathcal{C}_{\rm GW}}{\dd\ln\Delta}\\
 &\simeq -8.45
 \qquad (\Delta=\Delta_{\rm fid}) .
\end{aligned}
 \label{eq:width-mass-response}
\end{equation}
using the lognormal fit in Eq.~\eqref{eq:cgw-lognormal-fit}. Shifting the feature location gives \(\delta\ln f_0\simeq\delta N_{\rm pk}\). For fixed observed tensor properties and fixed shape,
\begin{equation}
\begin{aligned}
 \delta\ln M_\chi
 &\simeq -3\,\delta N_{\rm pk}
 +\frac12\delta\ln\mathcal{C}_{\rm GW}\\
 &\quad
 -\delta\ln\Delta
 -\frac92\delta(\Delta^2)
 -\delta\ln\Rthree .
\end{aligned}
 \label{eq:npk-shift}
\end{equation}
so depth, width, and location have separable effects in Fig.~\ref{fig:robustness}.

The fiducial background remains perturbative, produces a broad scalar enhancement, and sits far below the Gaussian PBH production threshold discussed next. The local ansatz is a representative source for the required \(\Pk(k)\). The moment closure applies equally to smooth peaks generated in monodromic-valley, dilaton-flattened, or anomaly-inspired inflationary settings \citep{Pirzada:2026jml,Pirzada:2026sle,Pirzada:2026uak}.

\section{Consistency conditions and theoretical robustness}
\label{sec:control}

\subsection{Reheating branch}

The radiation-era kernel in Eq.~\eqref{eq:sigw-master} applies when the scalar modes re-enter after reheating. The frequency associated with the end of reheating is
\begin{equation}
\begin{aligned}
 f_{\rm reh}&\simeq 2.65\times10^{-8}\,\mathrm{Hz}
 \left(\frac{T_{\rm reh}}{\mathrm{GeV}}\right)
 \left(\frac{g_*}{106.75}\right)^{1/2}\\
 &\quad\times
 \left(\frac{g_{*s}}{106.75}\right)^{-1/3}.
\end{aligned}
 \label{eq:rehfreq}
\end{equation}
which locates the transition between radiation-era and pre-radiation transfer functions \citep{Kofman:1997yn}.

For \(\fzero=2.979\,\mathrm{MHz}\), radiation-era re-entry requires
\begin{equation}
 T_{\rm reh}\gtrsim 1.12\times10^{14}\,\mathrm{GeV},
 \label{eq:rehreq}
\end{equation}
placing the benchmark on the prompt high-scale reheating branch. This branch is compatible with high-scale inflation when the vacuum tensor amplitude respects the current bound on \(r\) \citep{Tristram:2021tvh}.

Delayed reheating moves the relevant modes into an early matter-dominated or transition-era transfer problem. The radiation kernel is then replaced by the appropriate non-radiation kernel \citep{Kofman:1997yn,Inomata:2019ivs,Inomata:2019zqy}. We parametrize the shift by
\begin{widetext}
\begin{equation}
 h^2\Omega_{\rm GW}^{\rm pk}\big|_{\rm nonRD}
 =\mathcal{S}_{\rm reh}\!\left(\frac{f_0}{f_{\rm reh}},\Gamma_\phi/H\right)
 h^2\Omega_{\rm GW}^{\rm pk}\big|_{\rm RD},
 \qquad
 M_\chi^{\rm lim}\big|_{\rm nonRD}
 =
 \mathcal{S}_{\rm reh}^{1/2}
 M_\chi^{\rm lim}\big|_{\rm RD},
 \label{eq:reh-branch-shift}
\end{equation}
\end{widetext}
where \(\mathcal{S}_{\rm reh}\) depends on the reheating history. Continuous reheating often suppresses the induced signal through potential decay; sudden-transition idealizations can enhance it. The present benchmark uses the radiation branch, and low-reheating histories are separate transfer calculations.

\subsection{PBH tail}

The benchmark also has a tiny Gaussian PBH tail. We use this as a consistency diagnostic for the body of the spectrum, leaving the model-dependent non-Gaussian far tail explicit. In the Gaussian estimate,
\begin{equation}
 \sigma_\delta^2\simeq
 \frac{16}{81}\Pk^{\rm max},
 \label{eq:sigmadelta}
\end{equation}
and the Press--Schechter collapse fraction is
\begin{equation}
 \beta_{\rm PBH}\sim
 \frac{\sigma_\delta}
 {\delta_c\sqrt{2\pi}}
 \exp\left[
 -\frac{\delta_c^2}
 {2\sigma_\delta^2}
 \right].
 \label{eq:pbhbeta}
\end{equation}

For \(\Pk^{\rm max}=1.04\times10^{-3}\), Eq.~\eqref{eq:sigmadelta} gives \(\sigma_\delta=1.43\times10^{-2}\). The representative thresholds \(\delta_c=0.4,\ 0.45,\ 0.5\) give \(\log_{10}\beta_{\rm PBH}\simeq-171,\ -216,\ -266\). The benchmark is far below efficient PBH formation in the Gaussian estimate.

Non-Gaussian corrections can be written as
\begin{equation}
 \beta_{\rm PBH}^{\rm NG}
 =
 \mathcal{E}_{\rm NG}
 \beta_{\rm PBH}^{\rm G},
\end{equation}
where \(\mathcal{E}_{\rm NG}\) encodes the effect of higher-order statistics. Even at \(\delta_c=0.4\), reaching \(\beta_{\rm PBH}\sim10^{-20}\) would require \(\mathcal{E}_{\rm NG}\sim10^{151}\). This number quantifies the enormous separation between the benchmark peak height and the PBH-abundant tail.

Transient non-attractor evolution can generate localized non-Gaussianity \citep{Atal:2018neu,Atal:2019erb,Taoso:2021uvl}. The nonlinear relation between \(\zeta\) and the density contrast also modifies collapse probabilities \citep{Maldacena:2002vr,DeLuca:2019qsy,Young:2019yug}. Broad peaks, nonspherical collapse, and profile dependence further affect PBH constraints \citep{SatoPolito:2019hws,Musco:2020jjb}. Recent analyses sharpen the same point for small-scale power spectra and PBH limits \citep{Fujita:2025dip,Carr:2026pbh,Kushwaha:2026constraints}. For that reason, PBHs are used here as a tail check; the normalization comes from \(\Omega_\chi h^2\) and the low moments of \(\Pk\).

\subsection{Dark-sector history}

The dark sector also satisfies radiation and stability checks. For the benchmark,
\begin{equation}
 \Delta \Neff^{\rm GW}=
 \frac{8}{7}\left(\frac{11}{4}\right)^{4/3}
 \frac{1}{\Omega_{\gamma,0}}
 \int \dd\ln f\,\OmGW(f)
 \ll 10^{-10},
 \label{eq:neffgw}
\end{equation}
well below current and forecasted sensitivity to extra radiation \citep{Aghanim:2018eyx,Goldstein:2026neff}. The relativistic fermion stage gives \(\Delta N_{{\rm eff},\chi}\simeq2.7\times10^{-21}\).

A minimal completion is
\begin{equation}
 \mathcal{L}_{\rm dark}
 = i\bar\chi\gamma^\mu\partial_\mu\chi
 -y_S S\bar\chi\chi
 +\frac12(\partial S)^2
 -V(S)
 -\lambda_{HS}S^2H^\dagger H ,
 \label{eq:darklag}
\end{equation}
with an exact dark \(Z_2\) under which \(\chi\) is odd and \(S\) is even. During production, \(\langle S\rangle=0\), and \(\chi\) is effectively massless and conformal. Later, \(S\) relaxes to \(v_S\), generating \(M_\chi=y_Sv_S\). Since this occurs after production and before matter-radiation equality, it converts the conserved \(n_\chi a^3\) into nonrelativistic matter.

The hidden sector stays nonthermal when the portal rate remains below the Hubble rate,
\(
\Gamma_{HS}\sim\lambda_{HS}^2T
<
H\simeq1.66\sqrt{g_*}\,T^2/M_{\rm Pl}.
\)
The exact discrete gauge symmetry forbids Planck-suppressed decay operators for the superheavy fermion.

The mass-generation transition can avoid an independent stochastic background. A biased crossover potential,
\begin{equation}
 V_T(S)
 =
 \frac12\left(c_ST^2-m_S^2\right)S^2
 +\frac{\lambda_S}{4}S^4
 -\eta_S^3S
 +V_0,
 \label{eq:darkthermalpotential}
\end{equation}
has a minimum obeying
\begin{equation}
 \lambda_Sv_S^3
 +\left(c_ST^2-m_S^2\right)v_S
 -\eta_S^3
 =0 .
 \label{eq:vs-continuous}
\end{equation}
The minimum moves continuously with temperature; bubble nucleation, runaway walls, and a phase-transition gravitational-wave component are absent. The benchmark stochastic signal is then the scalar-induced background from the primordial curvature peak.

A first-order dark transition would add a separate source. If the released vacuum energy is
\begin{equation}
 \alpha_S
 \equiv
 \frac{\Delta V_S}{\rho_{\rm rad}(T_S)}
 \ll1,
 \label{eq:alphaS}
\end{equation}
then the standard phase-transition estimate gives
\begin{equation}
 \Omega_{\rm PT}^{\rm pk}h^2
 \lesssim
 10^{-6}\,
 \kappa_S^2
 \left(\frac{H_S}{\beta_S}\right)^2
 \left(\frac{\alpha_S}{1+\alpha_S}\right)^2 .
 \label{eq:ptgw-bound}
\end{equation}
Keeping this below the benchmark \(h^2\Omega_{\rm GW}^{\rm pk}\simeq5\times10^{-12}\) is guaranteed by
\begin{equation}
 \kappa_S\alpha_S
 \left(\frac{H_S}{\beta_S}\right)
 \ll
 2\times10^{-3}.
 \label{eq:ptgw-sufficient}
\end{equation}
This scaling follows standard first-order-transition results \citep{Caprini:2019egz,Ai:2025gtr}. The main analysis uses the crossover branch of Eqs.~\eqref{eq:darkthermalpotential}--\eqref{eq:vs-continuous}; a first-order branch is an additional source model beyond the benchmark.

\begin{figure*}[tbp]
\centering
\includegraphics[width=0.95\textwidth]{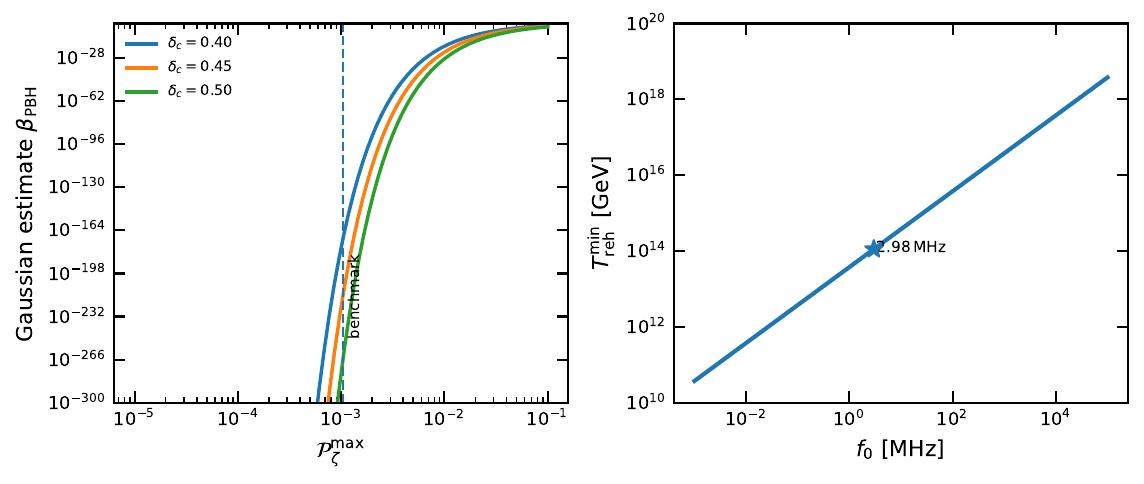}
\caption{Control checks. Left: Gaussian PBH tail estimate as a function of scalar peak height, with the benchmark marked; the text treats non-Gaussian PBH tails as a separate model-dependent diagnostic. Right: minimum reheating temperature required for the scalar scale to re-enter during radiation domination. The MHz benchmark sits on the high-reheating branch; low-reheating histories require a different transfer kernel.}
\label{fig:pbh-reheating}
\end{figure*}

\section{Robustness and reach}
\label{sec:robustness}

\begin{figure*}[tbp]
\centering
\includegraphics[width=0.95\textwidth]{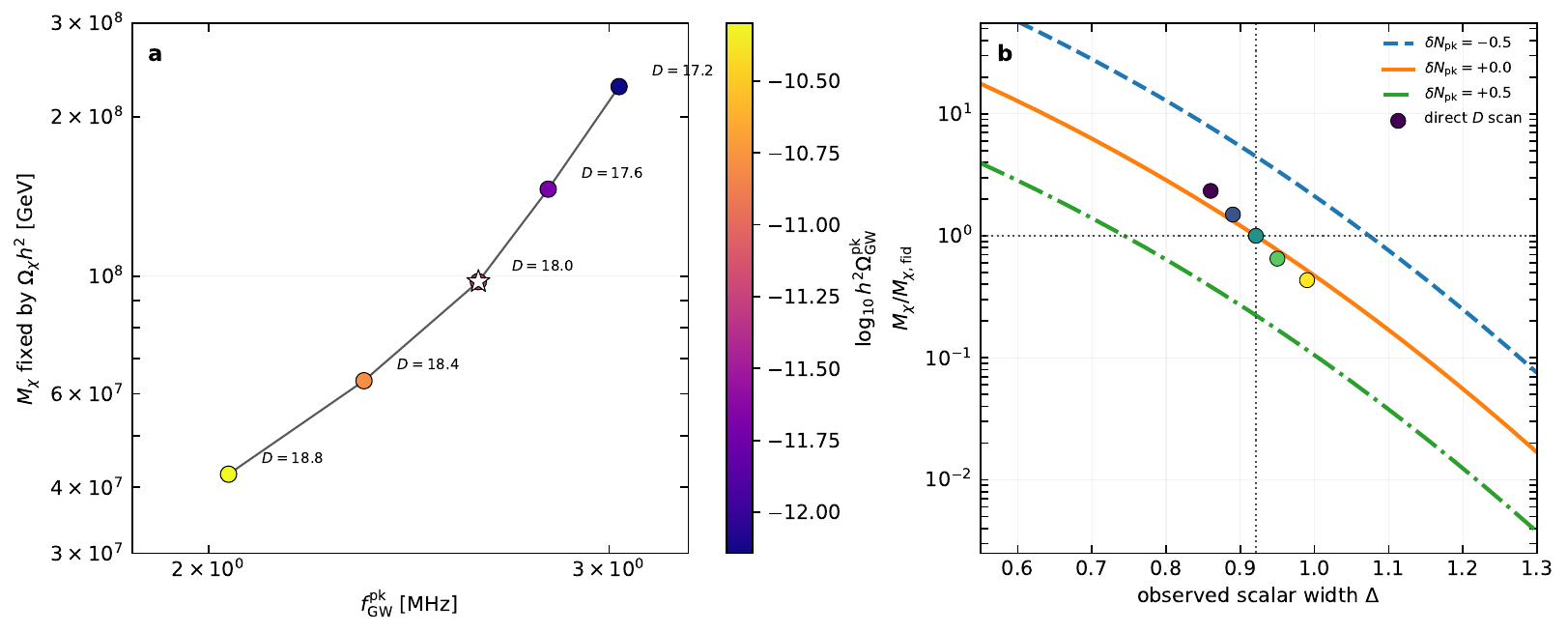}
\caption{Response to controlled feature deformations. Left: direct Mukhanov--Sasaki/SIGW family generated by varying \(D\), showing the abundance-fixed fermion mass against the induced-wave peak frequency and coloring by the tensor peak. Right: inferred-mass response to width \(\Delta\) and peak-location shift \(\delta N_{\rm pk}\); points show the direct \(D\) scan.}
\label{fig:robustness}
\end{figure*}

We test the closure relation by deforming the inflationary feature and recomputing every step. Figure~\ref{fig:robustness} varies the transient depth through \(D=17.2,\,17.6,\,18.0,\,18.4,\,18.8\). For each point we solve the Mukhanov--Sasaki equation, evaluate \(\Rthree\), impose \(\Omega_\chi h^2=0.12\), and perform the full radiation-era SIGW convolution. Larger \(D\) increases \(\Az\), lowers the abundance-fixed \(M_\chi\), and raises \(h^2\Omega_{\rm GW}^{\rm pk}\). The strongest point in this family, \(D=18.8\), gives \(\fzero=2.60\,\mathrm{MHz}\), \(f_{\rm GW}^{\rm pk}=2.04\,\mathrm{MHz}\), \(M_\chi=4.23\times10^7\,\mathrm{GeV}\), and \(h^2\Omega_{\rm GW}^{\rm pk}=5.01\times10^{-11}\).

Width and location probe different directions. Varying \(\sigma\) changes the measured \(\Delta\), while changing \(N_{\rm pk}\) shifts the characteristic frequency. Equation~\eqref{eq:width-mass-response} gives the leading mass response to width; Eq.~\eqref{eq:npk-shift} gives the frequency response. The direct \(D\)-scan, together with the analytic \((\Delta,N_{\rm pk})\) derivatives, identifies how a detected spectrum could fail the predicted width or mass--frequency relation.

\begin{figure*}[tbp]
\centering
\includegraphics[width=0.95\textwidth]{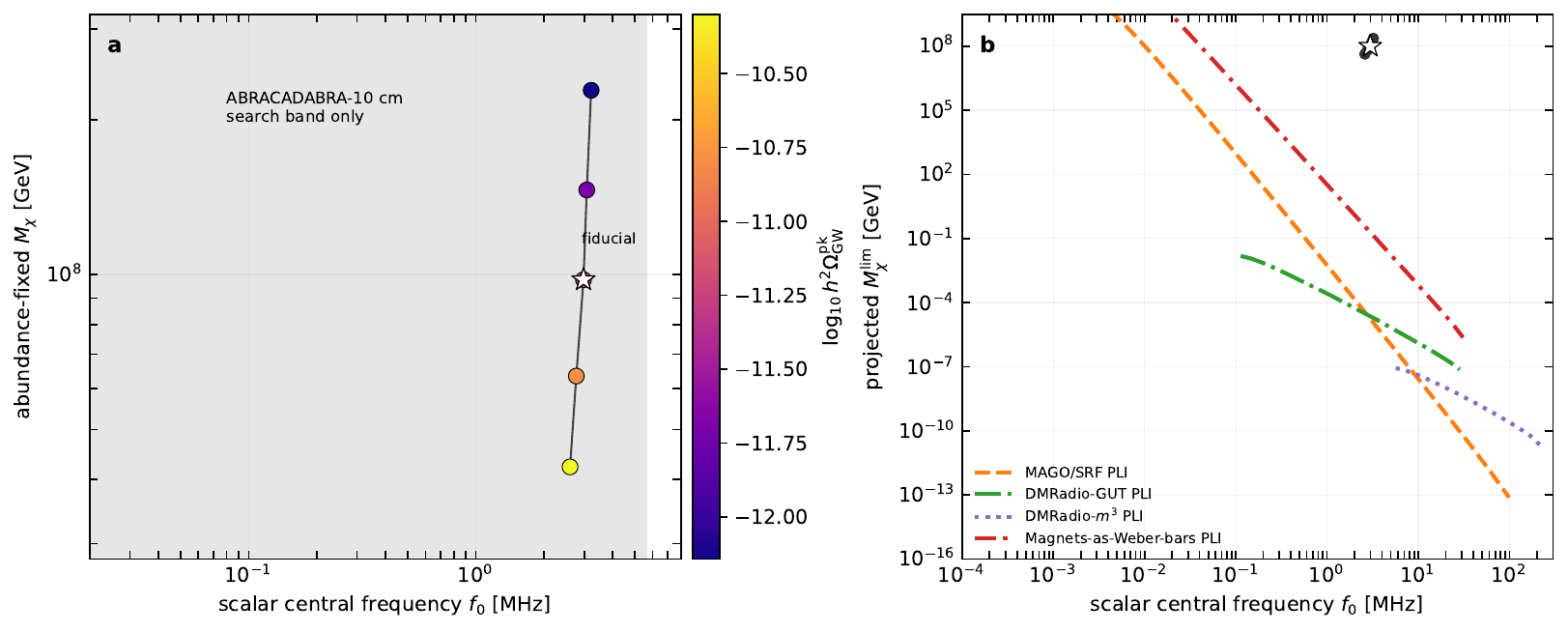}
\caption{Current search band and projected reach. Left: ABRACADABRA-10 cm is shown as a published search interval overlapping the controlled family; the panel treats it as search coverage and leaves benchmark-shape mass exclusions to likelihood analyses. Right: representative PLI projections translated with Eq.~\eqref{eq:massreach}. Separating the panels keeps published coverage distinct from projected reach indicators.}
\label{fig:mass-reach}
\end{figure*}

\begin{figure*}[tbp]
\centering
\includegraphics[width=0.98\textwidth]{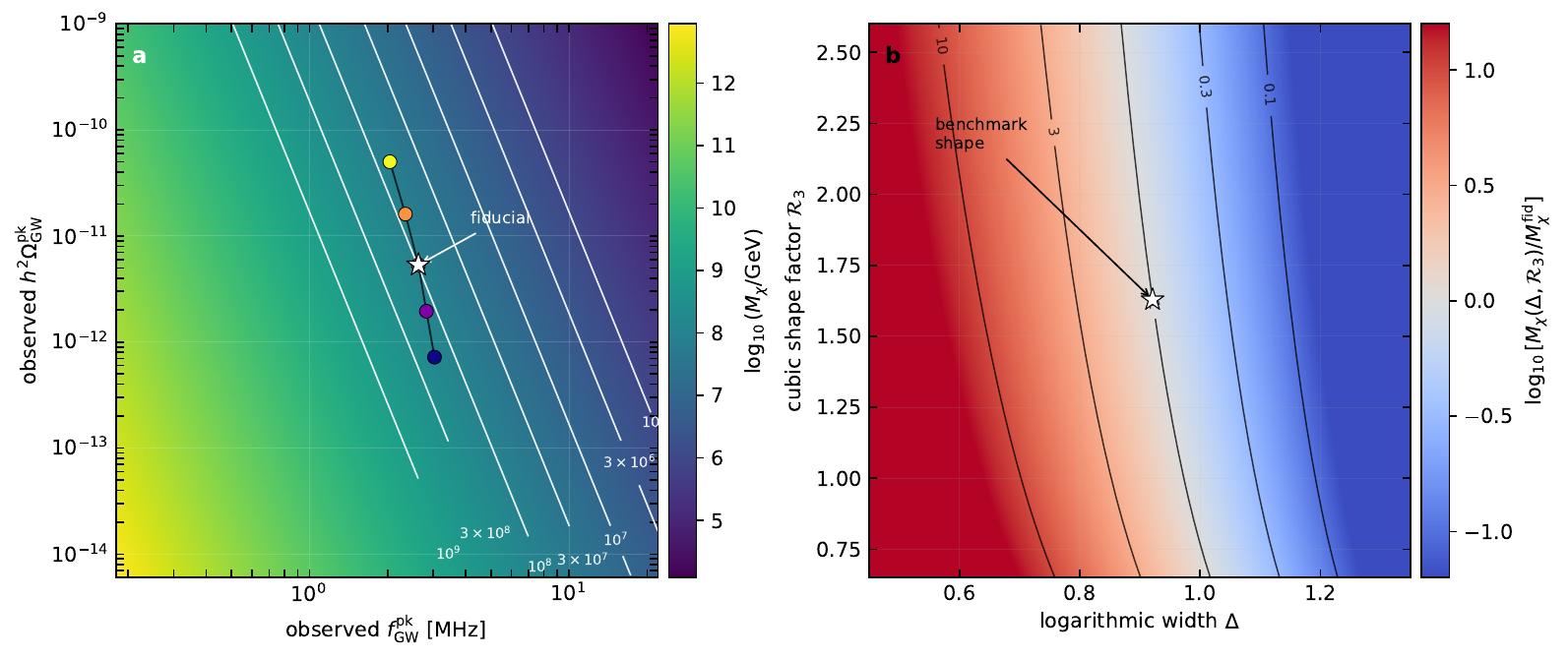}
\caption{Closure diagnostic for a stochastic MHz detection. Left: inferred conformal-fermion mass from Eq.~\eqref{eq:mass-inference} in the observed \((f_{\rm GW}^{\rm pk},h^2\Omega_{\rm GW}^{\rm pk})\) plane, with controlled-family points overlaid. The white star marks the fiducial direct-kernel benchmark. Right: mass shift for a fixed fiducial observed signal after varying \(\Delta\) and \(\Rthree\), including the fitted \(\mathcal{C}_{\rm GW}^{\rm LN}(\Delta)\). Contours give the multiplicative mass ratio; consistency fixes the mass, width, cubic factor, and quadratic shape factor together.}
\label{fig:closure-diagnostic}
\end{figure*}

Figures~\ref{fig:mass-reach} and~\ref{fig:closure-diagnostic} show the experimental consequences. Since \(h^2\Omega_{\rm GW}^{\rm pk}\propto\fzero^{-6}\), the mass reach falls rapidly with increasing frequency, making the low-MHz interval the most favorable part of the benchmark band. A null search sets a lower bound on \(M_\chi\) after the peak shape is specified. A detection has less freedom than a generic stochastic template: \(M_\chi\), \(\fzero\), \(\Delta\), \(\Rthree\), and \(\mathcal{C}_{\rm GW}\) all have to match one curvature spectrum.

\section{Conclusions}

A localized small-scale enhancement of \(\Pk\) can be tested through two low moments instead of through a PBH tail alone. In the scenario studied here, scalar inhomogeneities produce a stable conformal fermion through the cubic moment, while the same perturbations generate scalar-induced gravitational waves through a quadratic convolution. The observed relic density fixes the scalar area, so the MHz tensor amplitude loses the freely adjustable normalization of ordinary SIGW templates.

The calculation keeps the shape dependence explicit. The coefficient \(\Rthree\) measures the cubic fermion moment, and \(\mathcal{C}_{\rm GW}\) measures the quadratic tensor convolution. For the benchmark Mukhanov--Sasaki spectrum, \(\Rthree=1.6287\), \(\mathcal{C}_{\rm GW}=1.2665\times10^{-6}\), \(f_{\rm GW}^{\rm pk}=2.6275\,\mathrm{MHz}\), and \(h^2\Omega_{\rm GW}^{\rm pk}=5.3325\times10^{-12}\) after imposing \(\Omega_\chi h^2=0.12\). Controlled deformations in \(D\), \(\Delta\), and \(N_{\rm pk}\) show how the inferred mass changes with peak height, width, and location.

The benchmark assumptions are explicit. The MHz modes re-enter during radiation domination for \(T_{\rm reh}\gtrsim1.12\times10^{14}\,\mathrm{GeV}\); delayed reheating changes the transfer kernel through \(\mathcal{S}_{\rm reh}\). The Gaussian PBH estimate gives \(\log_{10}\beta_{\rm PBH}\simeq-171\) even for \(\delta_c=0.4\), while model-dependent non-Gaussian far tails remain a separate diagnostic. A crossover in the dark scalar sector generates \(M_\chi=y_Sv_S\) with no phase-transition gravitational-wave signal.

The experimental reading is direct. Equation~\eqref{eq:massreach} converts HFGW sensitivity into a lower bound on \(M_\chi\); Eq.~\eqref{eq:mass-inference} converts a measured stochastic peak into an inferred fermion mass. A credible MHz signal would be overconstrained by the relic abundance, tensor amplitude, peak frequency, spectral width, and shape factors. Cross-correlating distinct detector concepts can test that overconstraint because independent response functions measure the same \(\Delta\)-dependent spectral width.

This low-moment closure gives MHz gravitational-wave searches a concrete small-scale inflation observable tied to dark matter. It works below the PBH-abundant range and fixes the tensor amplitude through the relic density. The measurement to watch is the joint consistency of \(\Omega_\chi h^2\), \(f_{\rm GW}^{\rm pk}\), \(h^2\Omega_{\rm GW}^{\rm pk}\), \(\Delta\), and \(\Rthree\) for one primordial peak.

\bibliographystyle{unsrtnat}
\bibliography{refs}

\end{document}